\documentclass[a4paper,12pt]{article}
\usepackage{epsfig}
\usepackage{color}
\usepackage{hyperref, url}
\usepackage{fullpage}
\newcommand{\Fig}[1]{Fig.\ \ref{#1}}
\author{Arindam Mazumdar \\
\sl{Theory Division, Saha Institute of Nuclear Physics} \\
\sl{Bidhannagar-1/AF, Kolkata, India}}
\date{}
\begin{document}

\title{\bf Parameters of SUSY-Hybrid Inflation and the Process of Preheating}
\maketitle

\begin{abstract}
Process of preheating for supersymmetric hybrid model of inflation is generally analyzed in two different ways known as parametric resonance and tachyonic preheating. In a common frame-work, we study both the processes from the decay or growth of homogeneous inflaton field and inhomogeneous waterfall field. We find that these two processes in SUSY hybrid F-term inflation are not parameter independent; rather one process will be more preferable than the other depending on the values of parameters and Fourier mode of the waterfall field.  Parameters of the inflationary potential are constrained from the observed CMB data and these constrains help us to identify the process of preheating responsible for this model.
\end{abstract}


\section{Introduction}
Theory of inflation has been successful so far in describing the origin of observed cosmic microwave background(CMB) anisotropies.
Since last decade COBE and WMAP has reached a certain level in measuring the values of the cosmological parameters. Planck experiment is aimed at increasing the accuracy of those data in recent future. As the accuracy of measurement increased some models of inflation have been ruled out, and some models became more successful.

Standard hybrid inflation\cite{Linde:1993cn} has been ruled out just after the release of WMAP first year data\cite{Peiris:2003ff}. But supersymmetric version of hybrid inflation \cite{Dvali:1994ms} survived because it gives a red spectrum as observed by WMAP. At the same time this model suggests a negligible amplitude of gravity wave which is to be verified in upcoming Planck experiment\cite{Planck}.     

Preheating is a process of rapid decay of inflaton's energy to other forms just after end of inflation. There are different mechanisms of preheating. Firstly there is the old and standard concept of preheating by broad parametric resonance  \cite{GarciaBellido:1997wm}. Possibility of this mechanism to be efficient has been studied in the context of susy-hybrid model \cite{BasteroGil:1999fz}. Another process of preheating is called tachyonic preheating, which became more important than this standard one in susy-hybrid models because of the negative curvature of the potential after the end of inflation.

Analysis of preheating in susy hybrid model has been an important field of study since last one decade.  For preheating process in hybrid inflation \cite{GarciaBellido:1997wm}, the trajectory of the fields were analyzed taking the inflaton field and the waterfall field as homogeneous. Only small inhomogeneous perturbation was considered in the background of a large homogeneous component of the waterfall field. Parametric resonance was analyzed, taking fully inhomogeneous waterfall field  in ref.\cite{Zlatev:1997vd}.
 When tachyonic preheating is analyzed, waterfall field is generally taken to be completely inhomogeneous. In ref.\cite{BasteroGil:1999fz} for SUSY hybrid inflation an effectively one field potential was derived by assuming a trajectory of the homogeneous fields and neglecting the oscillation of inflaton. That form of potential was used in ref.\cite{Felder:2001kt} to analyze the tachyonic preheating process in susy-hybrid inflation case. Important numerical simulations have been done with LATTICEEASY\cite{Felder:2000hq} and DEFROST\cite{Frolov:2008hy} code. These codes helped greatly to study the growth of the perturbations with time. But often hard to identify the process responsible for this growth.   

 In this work we did not assume any effective form of the susy-hybrid potential. Rather we have taken the full two-field potential. Inflaton is considered completely homogeneous, but waterfall field is considered fully inhomogeneous, i.e. no background homogeneous component is assumed. So we do observe an effect of oscillating potential coming from the oscillation of inflaton field. Also, the negative curvature part in the potential boosts tachyonic growth of the modes of waterfall field. We will show here that in certain range of parameter space this oscillating potential will lead to efficient parametric resonance for some Fourier modes of waterfall field, while others modes may experience tachyonic growth. So, tachyonic preheating and parametric resonance are not two different processes to happen at two different times; rather at the same time different Fourier modes of waterfall field may encounter different processes. But for other ranges of parameters where amplitude of inflaton's oscillation decays down very quickly, parametric resonance is not so efficient. Therefore tachyonic preheating will be the main dominating process for most of the modes in that parameter range.  

So, firstly from the results of our analysis we can identify two different processes of preheating at the same time. Secondly, we do show that preheating process in susy-hybrid inflation is a parameter dependent process, not as it was expected in          
ref.\cite{BasteroGil:1999fz}. Depending on the values of parameter the process of preheating for different modes of the waterfall field gets changed.

 We will be focusing on the F-term susy-hybrid inflation in our work. Inflation occurs due to the 1-loop correction in the potential along certain flat direction. In every model of inflation some amount of parameter tuning is necessary to fit the observed data. So here also certain values of the parameters are favored for latest CMB observation. Can those tuned parameter say anything about the mechanism of preheating in susy-hybrid model?   In this 
report we will try to address this question.
  
 In first section we will first describe the SUSY hybrid inflation formalism and show how to calculate the observed CMB quantities out of it. After that different mechanisms of preheating will be discussed for this model. First we will describe the process of parametric resonance with fully inhomogeneous waterfall field. We will show that how different modes of the waterfall filed behave differently under parametric resonance. Condition for efficient parametric resonance will also be discussed in this section. Next we will briefly outline the tachyonic preheating process. It is mostly impossible to workout the fields' amplitude and occupation number analytically. So the analysis will be done mostly numerically. 

We present here a common formalism to describe both the processes simultaneously. Growth of occupation number  of different modes of the waterfall field gives the hints of what process is dominating at that situation. There is a technical problem of considering Hubble parameter ($H(t)$) in solving the equation of motion of these fields. Hubble parameter depends on the spatial average of the inhomogeneous filed. In mean field approximation equation of motion also involves the spatial average of waterfall field. So to solve this exactly one need to solve infinite number of coupled equations. This technical problem forces many authors to consider both fields as homogeneous. But here to overcome this problem we assume that the spatial average of the square of the inhomogeneous waterfall field initially grows exponentially as tachyonic preheating and settles at the minima of the potential. Assuming this we see the evolution of Hubble parameter. And then using that we solve the different modes of the waterfall field. After that we will show how different choices of parameters of the model can affect the process of preheating. We will also show the mode by mode pattern of the growth in occupation number for different parameter choice. We find a clear distinction between the patterns of the growth of occupation number for different choice of parameters.

 In the last section we will find out what values of observed CMB quantities can this model give to us. To fit with the observed values of these quantities parameters of the model will be tuned. Then we show what can those required values of parameters tell us about the mechanism of preheating. Choice of values of these parameters depends on the choice of 1-loop correction in the effective potential. There are lots of different models in the susy-hybrid class where different forms of one loop correction are used. Some of them have observational motivation and some of them have theoretical motivation. Some times extra symmetries are also imposed in the superpotential. For these different varieties of models tuned parameters will be different. So depending on that conclusion may differ from model to model. Here we take the old and standard susy-hybrid 1-loop correction just as an example. Although the analysis of preheating process does not depend on the 1-loop potential. It is because this effective potential is valid only in the time of inflation. After the end of inflation a soft susy breaking mass term drives inflaton to the minima, and the value of this term is very small compared to the 1-loop correction term. So our analysis of preheating process is valid for all F-term susy hybrid model. For different form of loop correction term just different values of parameters have to be chosen and from that exact preheating process can be immediately identified from our analysis.            
\section{General Formalism of SUSY Hybrid Inflation} 
SUSY hybrid inflation models are of two types, F-term inflation and D-term inflation.
F-term inflation attracts more interest than D-term inflation because it is tailor made to fit with Higgs' mechanism. 
In general F-term inflation assumes a superpotential of the following form\cite{{Dvali:1994ms}}.  
\begin{eqnarray}
 W=\kappa \Phi(\Psi\bar\Psi-M^2)
\end{eqnarray}
Here $\Psi,\bar\Psi$ are chiral scalar multiplets conjugate to each other. $\Phi$ is a gauge singlet scalar and contains inflaton ($\phi$).
Corresponding F-term scalar potential is as follows:
\begin{eqnarray}
 V_F=\kappa^2 (|\psi|^2-M^2)^2+2\kappa^2 |\phi|^2|\psi|^2 \label{hybridpot}
\end{eqnarray}
where $\psi$ and $\bar\psi$ can be any scalar multiplet conjugate to each other. In the context of applying grand unified theories(GUT) $\psi$ is thought to be the higgs of the corresponding GUT group. 
Inflation occurs in $\psi=\bar\psi=0$ trajectory. This is a flat direction with a constant potential of $\kappa^2M^4$. 
One loop correction in this direction gives rise to Coleman-Weinberg potential\cite{Coleman:1973jx} of the following form\cite{Huq:1975ue}:
\begin{eqnarray}
v(\phi)= {1\over 64\pi^2}\sum_{i}(-1)^{F_i}M_i^4\log{M_i^2\over\Lambda^2}
\end{eqnarray}
Here $F_i$ is the fermion number, $M_i$ is the mass eigen value of the component of $\Psi$ and $\Lambda$ is the renormalization scale.
 Total effective potential by which inflation is driven, turns out to be
\begin{eqnarray}
V(\phi) & = & \kappa^2M^4 + {\kappa^4{\cal N}\over 32\pi^2}\left( (\phi^2+M^2)^2\log{\kappa^2(\phi^2+M^2)\over\Lambda^2}\right. \\
&  &
+ \left. (\phi^2-M^2)^2\log{\kappa^2(\phi^2-M^2)\over\Lambda^2}- 2\phi^4\log{\kappa^2\phi^2\over\Lambda^2}
\right)\nonumber
\end{eqnarray}
where ${\cal N}$ is the dimensionality of the representation of $\psi,\bar\psi$. In total the potential has two minina, one is the local minima at $\phi=0,\psi=0$ and another is the global minima at $\phi=0, \psi=M$. Inflaton $\phi$ slowly rolls down through this effective potential during the period of inflation towards the local minima. This effective potenial is valid only in the range where $\phi> M$. After the end of inflation another susy-breaking term is required to drive inflaton to $\phi=0$. In general an extra mass term in the potential, say $m^2\phi^2$ is assumed for this purpose. But the value of $m^2$ has to be smaller than the co-efficient of $\phi^2$ arising from the loop-correction $\kappa^4 M^2$. Otherwise this term will effect the values of the observed CMB quantities at the time of horizon-exit of a particular mode of inflaton. 
     
Therefore, in general these types of potentials can be divided into two parts  
\begin{eqnarray}
 V(\phi)= V_0 +v (\phi).
\end{eqnarray}
where $V_0$ is the constant vacuum energy density and    
$v(\phi)$ is the loop correction term.  

Observed CMB anisotropy parameters are mainly scalar spectral index($n_s$) and
amplitude of density perturbation ($\Delta_{\mathcal{R}}$). These quantities are related to the inflation potential
through the slowroll parameters($\epsilon, \eta$). 
\begin{eqnarray}
n_s & = & 1-6\epsilon +2\eta\\
 \Delta^2_{\mathcal{R}} & = & {8\over m^4_{Pl}}\left.{V\over \epsilon}\right|_{k=aH} 
\end{eqnarray}
where,
\begin{eqnarray}
 \epsilon & = & {m^2_{Pl}\over 16\pi} \left({V_{\phi}\over V}\right)^2 = {m^2_{Pl}\over 16\pi} \left({v_{\phi}\over V_0+v}\right)^2, \\
\eta & = & m^2_{Pl} \left({V_{\phi\phi}\over V}\right)= m^2_{Pl} \left({v_{\phi\phi}\over V_0+v}\right).
\end{eqnarray}
Here $\phi$ in subscript means derivative with respect to $\phi$ and $m_{Pl}=1/\sqrt{G}$ and has a value of $1.22\times 10^{19}$ GeV.
 Amplitude of tensor perturbation  is related to inflation potential as 
\begin{eqnarray}
\Delta_h^{2}=\left.{128\over 3m_{Pl}^4}V(\phi)\right|_{k=aH}
\end{eqnarray}
So tensor-to-scalar ratio takes the form
$r= {\Delta_h^{2}\over \Delta^{2}_{\mathcal{R}}}=16 \epsilon $. 
It can be shown that the value of $v(\phi)$ is negligible compared to $V_0$, so 
\begin{eqnarray}
r\approx{128 \over 3 m_{Pl}^4 \Delta^{2}_{\mathcal{R}}}\kappa^2 M^4
\end{eqnarray}
SUSY hybrid model is a model with negative $\eta$, which gives negligibly small tensor-to-scalar ratio. Since no gravity wave signal has yet been observed in WMAP, it puts an upper bound on the tensor-to-scalar ratio ($r$)  as 0.36 \cite{Larson:2010gs}. If Planck is able to measure any $r$ with a value of $0.1$ or above this type of model will be discarded. But if it fails to detect $r$, it will also give an experimental restriction on the values of the parameters $\kappa, M$.  
   
 In standard hybrid inflation scenario $\phi$ reaches a critical value $\phi=\phi_c=M$, after
which waterfall occurs and gives rise to the end of inflation. But for this kind of 1-loop potential slowroll parameters becomes close to unity before $\phi$ reaches $M$ and inflation ends little earlier than the usual case. We have numerically found that a particular relation between $\kappa$ and $M$ will allow $\Delta^2_{\cal{R}}$ to be in the range of observed values.     



\section{Different Processes of Preheating }
Mechanism of preheating is a momentum dependent process, i.e. for inhomogeneous field different Fourier component evolves differently with time. In parametric resonance process\cite{Kofman:1994rk} for some values of the wavenumber resonance occurs, while for others values it may not. In tachyonic preheating \cite{Felder:2000hj} different modes experience different force term in the equation of motion depending on their wavenumber. In general both of them gives solution depending on wavenumber of the Fourier mode. But tachyonic preheating if encountered by a Fourier mode is lot more effective than parametric one. 

\subsection*{Parametric Resonance}     
 In our particular model inflation ends much earlier than $\phi$ reaches the critical value $M$. So for certain ranges of values of $M$ and $\kappa$ inflaton will sustain an oscillation around local minima $\phi=0, \psi=0$. 
Now we Fourier decompose the spatial part of $\psi$ as $\sum_{k}\psi_k(t)e^{ikx}$ where $k$ is the co-moving wavenumber.
As usual inflaton $\phi$ is regarded as spatially homogeneous. We derive the equations of motion for these two fields from equation(\ref{hybridpot}) and taking mean-field approximation we get 
\begin{eqnarray}
\ddot{\psi}_k+3H\dot\psi_k+ {k^2\over a^2}\psi_k+4\kappa^2\langle\psi^2\rangle\psi_k+4\kappa^2\phi^2\psi_k-4\kappa^2 M^2\psi_k & = & 0 \label{eom} \\
\ddot{\phi}+3H\dot\phi+2\kappa^2\langle\psi^2\rangle\phi+m^2\phi & = & 0 \label{eom2}
\end{eqnarray}   
$\langle \psi^2\rangle$ is the spatially averaged value of $\psi^2$.  By taking $X_k=a^{3/2}\psi_k$ we re-write equation(\ref{eom}) as:
\begin{eqnarray}
\ddot{X_k}+ \left[{k^2\over a^2}+4\kappa^2\phi^2+4\kappa^2{\langle X^2\rangle\over a^3}-4\kappa^2 M^2\right] X_k=0
\end{eqnarray}
Here we have neglected the pressure term since at the time of free oscillation it is vanishingly small. 
Frequency of oscillation of the $\phi$ field, $\omega_{\phi}$, can be estimated from (\ref{eom2}) at a particular moment as 
\begin{eqnarray}
\omega_{\phi}= \omega_0\sqrt{1-\left({3H\over 2\omega_0}\right)^2}\label{omega}
\end{eqnarray}
where $\omega_0^2= 2\kappa^2\langle\psi^2\rangle + m^2$. For some ranges of $M, \kappa$ values of $\omega_{\phi}$ is real and for some ranges it will be imaginary. This means that for the real values of $\omega_{\phi}$ we can get a sustainable oscillation in $\phi$ field and for imaginary values the oscillation will be over-damped, i.e. $\phi$ will quickly settle at the local minima. 

For oscillatory case of $\phi$ we assume for a certain period $\phi=\phi_0 \cos \omega_{\phi}t$ and change the variable $t$ to $\tau =\omega_{\phi}t$. This way we arrive at the familiar form of Mathieu equation\cite{GarciaBellido:1997wm} from (\ref{eom})
\begin{eqnarray}
{d^2 X_k\over d\tau^2}+[A_k+2q \cos(2\tau)]X_k=0\label{Mathieu}
\end{eqnarray} 
where 
\begin{eqnarray}
A_k & = &{k^2/a^2+2\kappa^2\phi_0^2+4\kappa^2{\langle X^2\rangle\over a^3}-4\kappa^2 M^2\over\omega^2_{\phi}} \\
q & = & {\kappa^2\phi_0^2 \over \omega^2_{\phi}}
\end{eqnarray}
Solution of Mathieu equation takes the form like $X_k\sim e^{\nu_k\tau}$ where $\nu_k$ is the critical exponent and function of $A_k$ and $q$. In an instability band of Mathieu chart $\nu$ gets a real value which gives exponential growth in $X_{k}$. Occupation number of $\psi_k$ can be calculated from the solution of eqn.(\ref{Mathieu}) or eqn.(\ref{eom}). Since eqn.(\ref{Mathieu}) assumes a purely oscillatory solution of $\phi$ we will use  numerical solution of eqn.(\ref{eom}) to get the occupation number from the well known formula
\begin{eqnarray}
n_k= {\omega_{k}\over 2}\left({|\dot{\psi_k}|^2\over\omega_k^2}+|{\psi_k}|^2-{1\over2} \right)\label{number}
\end{eqnarray}
where $\omega^2_k={k^2\over a^2}-4\kappa^2M^2$
and $n_k$ is proportional to $|X_k^2|\sim e^{2\nu_k\tau}$.
     
In an expanding universe parametric resonance  is always expected to be in broad resonance regime. Values of $A_k$ and $q$ both change with time and parametric resonance is not constrained in a single instability band of Mathieu chart. So by observing the growth of $n_k$ we can find the average critical exponent. In the $A,q$ parameter space broad parametric resonance region is identified to be in the area under  $A\leq 2q+\sqrt{q}$\cite{GarciaBellido:1997wm,Zlatev:1997vd}. This part of the parameter space can give efficient parametric resonance.  

\subsection*{Tachyonic Preheating}
 Tachyonic preheating is a process in which if a field has tachyonic instability around some extrema, its amplitude can increase explosively and preheating gets complete with a single oscillation of that field or even before a single oscillation. In our particular model as the inflaton $\phi$ crosses the critical value $M$, potential for $\psi$ becomes tachyonic. Amplitude of $\psi_k$ keeps increasing as long as $k^2+4\kappa^2\psi_k^2+4\kappa^2\phi^2-4\kappa^2 M^2$ remains negative. Spontaneous symmetry breaking gets completed after it reaches $M$. Occupation number of the $\psi_k$ can be defined as equation(\ref{number}). Since $\omega^2_k={k^2\over a^2}-4\kappa^2M^2$ is negative for ${k^2 \over a^2}< 4\kappa^2M^2$ we use $\omega^2_k={k^2 \over a^2}$ without any loss of generality \cite{Felder:2001kt}.
If initially $\psi_k$ and $\langle\psi^2\rangle$ is small compared $\phi$ and $M$, we can have  
\begin{eqnarray}
\psi_k\sim {\rm exp}(t\sqrt{4\kappa^2 M^2-k^2-4\kappa^2\phi^2})\label{growth}
\end{eqnarray}
Damping term is neglected for the time being. $\phi$ decreases with time and $\langle\psi^2\rangle$ increases very fast. So, within a short time growth of $\psi_k$ and $\langle\psi^2\rangle$ gets complete. 
In the same way initial growth of occupation number can also be estimated to be
\begin{eqnarray}
n_k\sim |\psi_k^2|\sim {\rm exp}(2t\sqrt{4\kappa^2 M^2-k^2-4\kappa^2\phi^2})
\end{eqnarray}

So far tachyonic preheating was analyzed neglecting the oscillating $\phi$ field. Here we take a different approach to analyze tachyonic preheating and parametric resonance in a single frame.  We use initial form of 
$\langle\psi^2\rangle$ as 
\begin{eqnarray}
\langle\psi^2\rangle & = & \int_{0}^{\sqrt{4\kappa^2\phi^2-4\kappa^2 M^2}} { dk^2\over 8\pi^2}{\rm exp}(2t\sqrt{4\kappa^2 M^2-k^2-4\kappa^2\phi^2})\\
& = & {1\over 16\pi^2 t^2}\left(1+ (2t\sqrt{4\kappa^2\phi^2-4\kappa^2 M^2}-1)e^{2t\sqrt{4\kappa^2\phi^2-4\kappa^2 M^2}}\right) 
\end{eqnarray}
Then after some time when $\langle\psi^2\rangle$ reaches $M^2$ we assume that it settles down at that value.  
 
Since Hubble parameter $(H(t))$ is also a function of the fields and their derivatives, this assumption allows us to see the effect of expansion of the universe at the time of preheating. Many authors used to assume static universe at the time of preheating to get read of this complicacy. Hubble parameter acts as a damping term in the equation of motions whose effect increases as the parameter $M$ and $\kappa$ increases in the potential. Physical wavenumber changes with  time as      
  \begin{eqnarray}
  {k\over a(t)}= {k\over {\rm exp}\displaystyle{\int_0^t H dt}} 
  \end{eqnarray}
At the time of the start of preheating $a$ has been taken to be 1. We did not assume any particular distribution of the initial amplitude of the modes. For all modes, initial amplitude is taken to be very small compared to $\phi$ and $M$. Our analysis can be done with any form of initial amplitude distribution. Although we find that initial amplitude cannot change the process of preheating; it can only change the time taken by the particular mode to start growing.        
  
\begin{figure}
\centering
\begin{tabular}{rr}
\epsfig{file=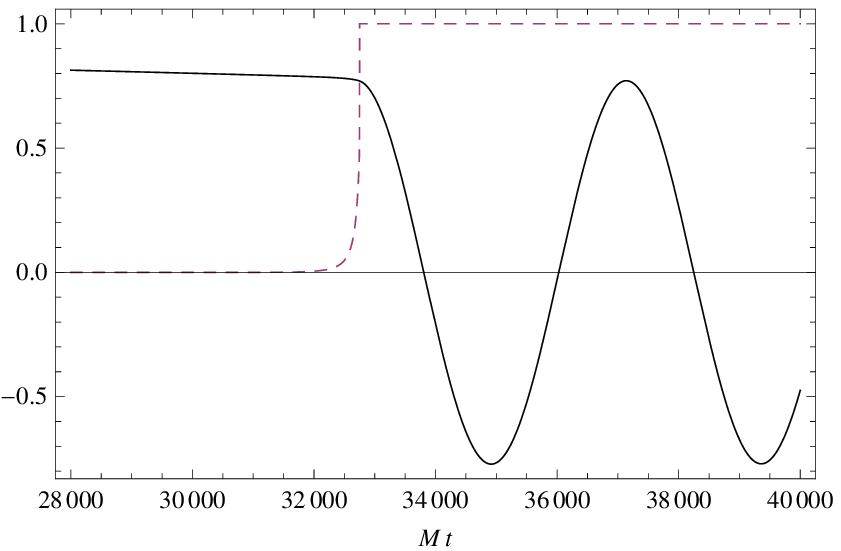,width=0.44\linewidth,clip=} &
\epsfig{file=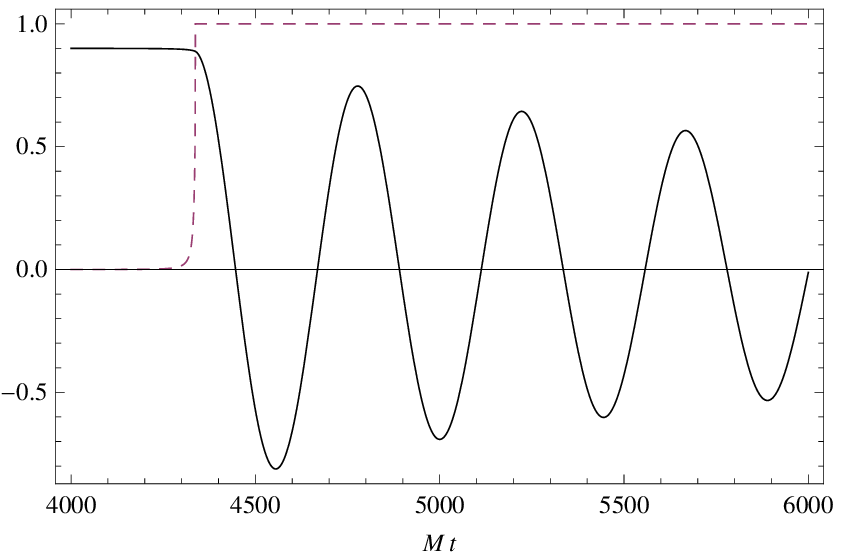,width=0.45\linewidth,clip=} \\
\epsfig{file=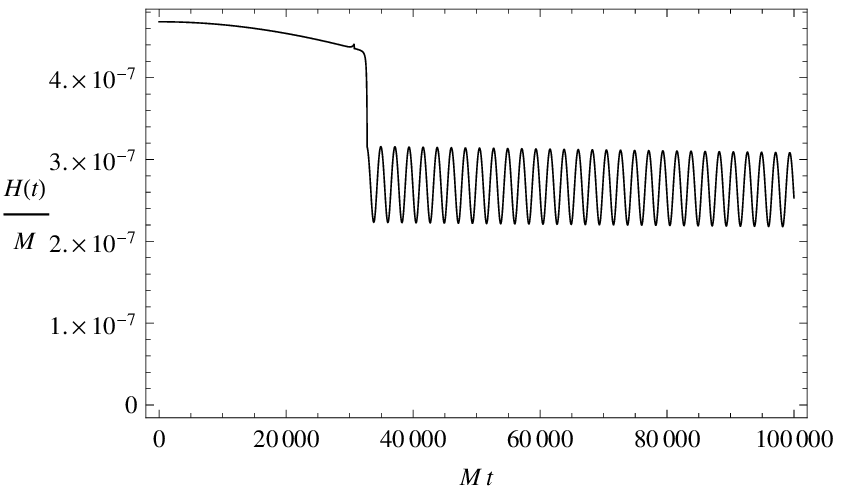,width=0.5\linewidth,clip=} &
\epsfig{file=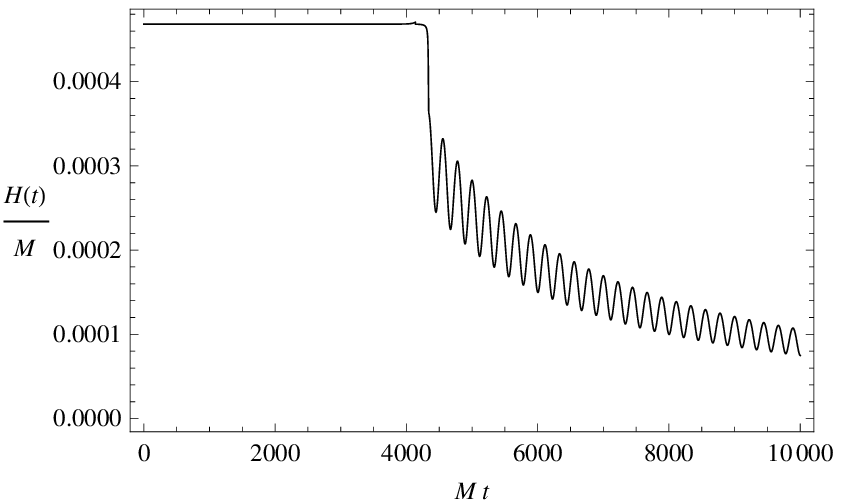,width=0.5\linewidth,clip=}\\
\end{tabular}
\caption{ $\phi(t)$ (solid line) and $\langle\psi^2\rangle$ (dashed line) are plotted in first row. Hubble parameter $H(t)$ is plotted in second row.  Values of the parameters are $\kappa=0.001$ and $M=1.22\times 10^{15}$GeV (left panel)  and $\kappa=0.01$ and $M=1.22\times 10^{17}$GeV (right panel)}
\label{fig1}
\end{figure}

\begin{figure}
\centering
\begin{tabular}{cc}
\epsfig{file=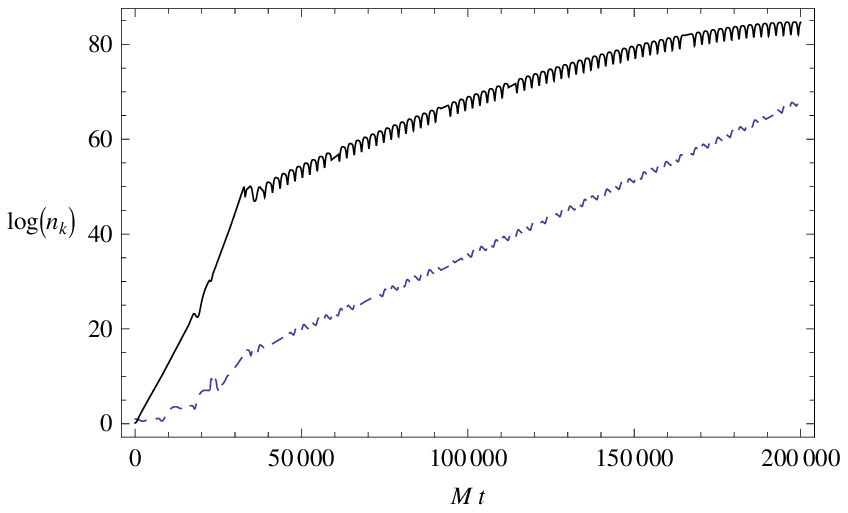,width=0.5\linewidth,clip=} &
\epsfig{file=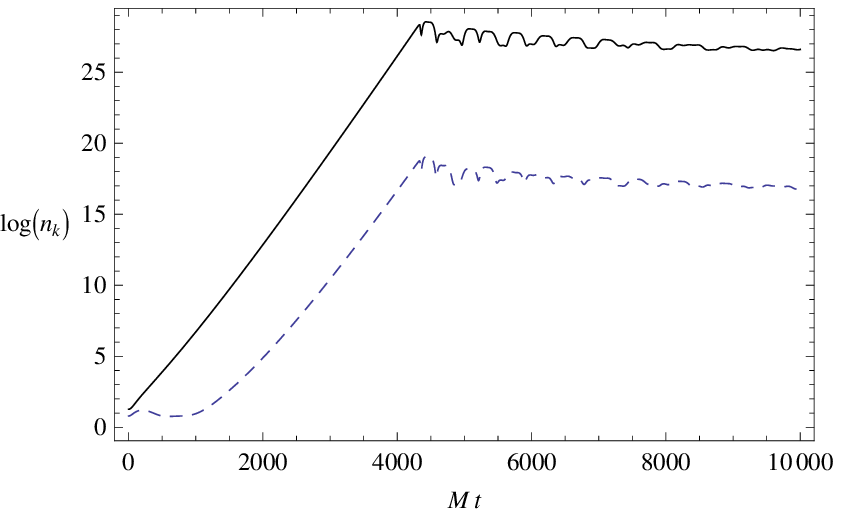,width=0.5\linewidth,clip=} \\
\end{tabular}
\caption{ Growth of occupation number $n_k$ for $k^2=0.1(4\kappa^2 M^2)$ (solid line) and $k^2=0.3(4\kappa^2 M^2)$ (dashed line) has been plotted with $\kappa=0.001$ and $M=1.22\times 10^{15}$GeV (left panel)  and $\kappa=0.01$ and $M=1.22\times 10^{17}$GeV (right panel)}
\label{fig2}
\end{figure}

  The left panel of \Fig{fig1} and \Fig{fig2} shows the plot of different quantities for $M=1.22\times 10^{15}$ GeV and $\kappa=0.001$ and right panel corresponds to  $M=1.22\times 10^{17}$ GeV and $\kappa=0.01$. The first row in  \Fig{fig1} shows how $\phi(t)$ and $\langle\psi^2\rangle$ behave with time. Next we solve the coupled equations(\ref{eom},\ref{eom2}) for Fourier component $\psi_k$ for $k^2=0.1(4\kappa^2 M^2)$ and $k^2=0.3(4\kappa^2 M^2)$ using the above form of $\langle\psi^2\rangle$. We find both the modes to oscillate around $\psi_k=0$ (not shown in the figure). This goes contrary to the standard expectation of parametric resonance that $\psi_k$ will oscillate around the global minima $\psi_k=M$. The reason behind it is the oscillating part in the potential coming from the oscillation of $\phi$. We see that among these two parameter sets for the first one damping effect is not so visible but for the second set of parameters $\phi$ decays rapidly due to the effect of high $H(t)$.
  
  In second row of \Fig{fig1} we plot $H(t)$ with time. This $H(t)$ is used in equations(\ref{eom},\ref{eom2}). We see due to the damping effect of the field $\phi(t)$ effects the $H(t)$ decreases with time rapidly for the second set of parameter, whereas for the first set of parameter average value of $H(t)$ remains almost constant.   
               
In \Fig{fig2} we plotted the growth of occupation number($n_k$) with time for two different co-moving wavenumbers. From the features of $\log(n_k)$ we can easily identify two different processes. The lower $k$ mode in the first set of parameters gets a steep increase initially. After the start of the oscillation of $\phi$ tachyonic preheating ends but        
parametric resonance becomes effective. For the lower $k$ mode in the same parameter set we do not see any significant tachyonic growth in $n_k$, rather the parametric resonance becomes effective from start of preheating. In the second parameter set we see that tachyonic preheating is effective in both $k$ modes. There is almost no effect of parametric resonance due to the high damping effect of large $H(t)$. $k^2=0.3(4\kappa^2 M^2)$ should not have undergone tachyonic preheating because initially it does not experience negative curvature term in the potential. But the due to large $H(t)$ physical wavenumber decreases very fast which gives rise to negative curvature term as well as tachyonic growth for this mode also.        
  
  We can analytically distinguish between the two processes from the log$(n_k)$ plot of both panels in \Fig{fig2}.  In the left panel we look at the $Mt$ range between 0 to 20000 for $k=0.1(4\kappa^2M^2)$. We know the amplitude of $\langle \psi^2\rangle$ is negligible compared to that of $\phi$ in this range. So curvature of the potential remains negative in this range for this particular mode. Therefore $\psi_k$ is in the form of equation(\ref{growth}). So the slope of log($n_k$) with respect to $Mt$ should come as $1.68\kappa$. From the plot we get this slope to be $1.25\times 10^{-3}$ in this $Mt$ range. So this result is close to the expected value. This assures that the process is tachyonic in nature. This tachyonic amplification should not be confused with the narrow parametric resonance shown for some model in Ref.\cite{GarciaBellido:1997wm}. Now we look at $k=0.3(4\kappa^2M^2)$ line of the same plot for $Mt$ range 0 to 30000. We find that curvature term for this mode is positive in this range. This gives an oscillating solution of $\psi_k$. So the gradual growth of $\log(n_k)$ is due to parametric resonance for this $k$. Therefore for these values of $M$ and $\kappa$ both the processes are present in different modes. 
  
  In right panel of \Fig{fig2} we do the same analysis and we find that for both values of $k$ tachyonic preheating is the main mechanism. Due to high damping factor the amplitude of $\psi_k$ gets so much damped that parametric resonance can not become effective. We know that the difference between $\log(n_k)$ for different values of $k$ can give rise to many interesting results like the parameters of non-Gaussianity and nonlinearity\cite{Enqvist:2004ey,Barnaby:2006cq,Barnaby:2006km}. In this analysis we see that this difference gets changed depending upon the parameters of the theory. Where in the left panel of \Fig{fig2} average slope of log$(n_k)$ is $1.66\times 10^{-3}$ and $0.5\times 10^{-3}$ for low and high $k$ at intial time, in right panel it is 0.007 and 0.004 for the corresponding $k$ values. 
  
There is a distinct difference between the time scale of the right panel and left panel. This comes from the frequency of the oscillation of $\phi$ field $\omega_{\phi}$ in equation(\ref{omega}). Since $\omega_0^2$ is proportional to $\kappa^2$ and $H^2$ is also proportional to $\kappa^2$ time period varies inversely with $\kappa$. This observation is also in accordance with ref.\cite{Felder:2001kt}, where time taken by the field to reach the zero curvature point in the potential was expected to vary as $1/\kappa$

For analyzing preheating in SUSY F-term hybrid model potential of the following form was assumed \cite{BasteroGil:1999fz} 
\begin{eqnarray}
V(\psi)=\kappa^2 M^2+3\kappa^2\psi^4-4\kappa^2\psi^3M \label{cubicpot}
\end{eqnarray} 
Potential in eqn.(\ref{hybridpot}) takes this form if we assume a trajectory for fields as $\psi=\pm(\phi-M)$. Growth of the perturbations with time had been shown for this potential using LATTICEEASY\cite{Felder:2001kt}. But different mechanisms are not unidentifiable from this work. In our analysis two different mechanisms are clearly identifiable.  

 

\section{Results and Discussion}

\begin{figure}
 \centering
\epsfig{file=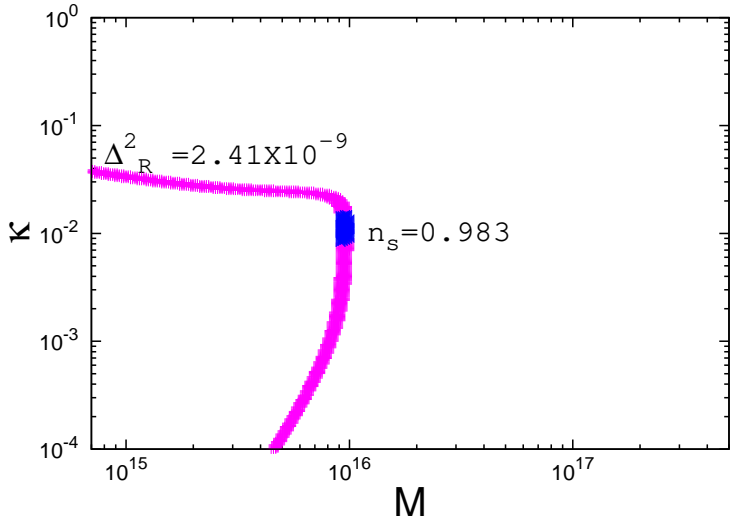,width= 4in} 
\caption{The curved line in $\kappa, M$ space belongs to the values of $\kappa$ and $M$ for which $\Delta^2_{\cal{R}}=(2.41\pm 0.11)\times 10^{-9}$. The blue part of the line corresponds to the $n_s=0.983$, the lowest value possible for the model.} 
\label{fig3}
\end{figure} 
We have shown in \Fig{fig3} that there is relation between $\kappa$ and $M$ for the particular observed value of $\Delta^2_{\cal R}$ which is $(2.41\pm 0.11)\times 10^{-9}$. To do the analysis we have taken the number of {\it e-foldings}, $N$, corresponding to the WMAP pivot scale 0.002 ${\rm MPc}^{-1}$, to be equal to 60. And $\cal N$ has been taken to be 8. For a particular value of $M$ there are two possible values of $\kappa$. So to choose the appropriate value of $\kappa$ and $M$ we have to look at the other observables. Points in this curve correspond to different values of $n_s$ and $r$. For all the points, the value of $r$ is of the order of $10^{-7}$, which is negligibly small as expected in negative $\eta$ models.  $n_s$ achieves its lowest value of 0.983 around $M=9.55\times 10^{15} $GeV and $  \kappa = 0.008-0.01$. This is a problem with the standard SUSY-hybrid model. This lowest value of $n_s$ is not in accordance with the latest measurement by WMAP, which gives $n_s=0.963\pm^{0.14}_{0.12}$.  
To fit with this observation various correction terms are added in this effective potential\cite{Rehman:2009yj,Mazumdar:2010sa}. In all those models for the observed value of $\Delta^2_{\cal{R}}$ a relation between $\kappa$ and $M$ can be found. 
As we have said earlier we chose the standard form of the susy-hybrid inflation just for an example.

\begin{figure}
\centering
\begin{tabular}{cc}
\epsfig{file=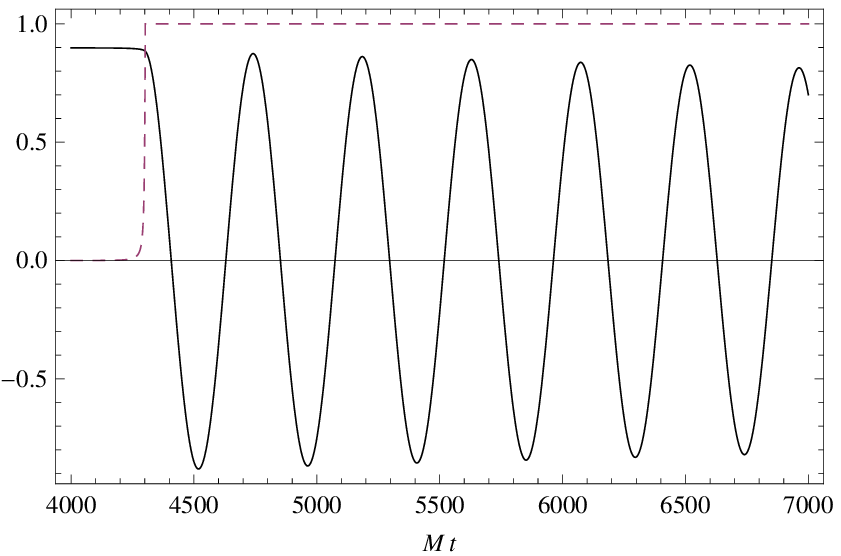,width=0.46\linewidth,clip=} &
\epsfig{file=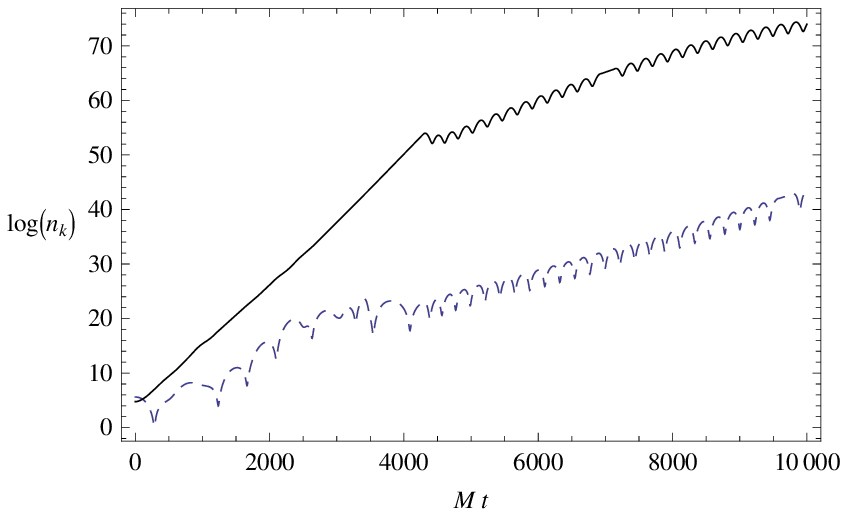,width=0.5\linewidth,clip=} \\
\end{tabular}
\caption{In left panel $\phi(t)$ (solid line) and $\langle\psi^2\rangle$ (dashed line) are plotted for $\kappa=0.01$ and $M=9.55\times 10^{15}$GeV. In right panel, for the same values of parameter, growth of occupation number $n_k$ has been plotted for $k^2=0.1(4\kappa^2 M^2)$(solid line) and $k^2=0.3(4\kappa^2 M^2)$(dashed line).}
\label{fig4}
\end{figure}

We have taken this tuned values of parameters in our analysis of preheating and the results are plotted in \Fig{fig4}.
We see that both the $k^2=0.1(4\kappa^2 M^2)$ and $k^2=0.3(4\kappa^2 M^2)$ behaves almost in the same fashion of $\kappa=0.001$ and $M=1.22\times 10^{15}$ GeV case. Lower $k$ mode undergoes tachyonic preheating initially and higher one undergoes parametric resonance. Average of the critical exponent $(\nu_k)_{\rm avg}$ for the higher $k$ mode has been estimated from log$(n_k)$ plot of this figure and  it comes to be 0.22.   
For the lower $k$ mode growth rate of log$(n_k)$ is 0.012 with respect to $Mt$. This parameter fixing is necessary for the cases where gravitational wave production is predicted from preheating \cite{Dufaux:2010cf}.  

Broad parametric resonance is generally thought to occur for the oscillation of $\phi$ around global minima. In that case value of $\psi$ is around $M$ and amplitude of $\phi$ is also of the order of $M$. Here we have seen that just after the end of inflation if inflaton can sustain oscillation around the global minima preheating via broad parametric resonance is possible. For higher values of $\kappa$ and $M$ amplitude of $\phi$ and $\psi_k$ decays down quickly which helps tachyonic preheating to dominate over parametric resonance. 

 Standard susy-hybrid model has only two parameters $\kappa$ and $M$. But other models may involve some more parameter. In that case it might be hard to pinpoint the values of parameters for which it will give two different preheating process. But this kind of models has been highly popular in the context of using Grand Unified Theories (GUT) in inflation. It was expected to get $M$ as the vacuum expectation value of higgs multiplet from the symmetry breaking scale of a particular GUT group. $\psi$ will have  the dimensionality of higgs. If those kinds of attempt are successful in future it will restrict the value of $M$. So, in those cases even with higher number of parameters it will be possible to tell about the exact preheating process. 
 
      Exact dynamics of preheating is necessary to understand for studying nonlinear evolution of density perturbation\cite{Frolov:2010sz}. Imprint of preheating can be observed in density perturbation of large-scale structure which originates from the super-horizon modes of fluctuation of the scalar fields at the end of inflation\cite{Chambers:2007se}. So, to find out the best possible model of inflation, along with reproducing other CMB observables, analysis of preheating process is also important, first step to which can be identifying the actual process.   

\paragraph{Acknowledgments: }We thank  Palash B Pal and Koushik Dutta for discussions and valuable
suggestions. 




\begin{thebibliography}{99}
\bibitem{Linde:1993cn} 
  A.~D.~Linde,
  Phys.\ Rev.\ D {\bf 49}, 748 (1994)
  [astro-ph/9307002].
\bibitem{Peiris:2003ff} 
  H.~V.~Peiris {\it et al.}  [WMAP Collaboration],
  Astrophys.\ J.\ Suppl.\  {\bf 148}, 213 (2003)
  [astro-ph/0302225].
\bibitem{Dvali:1994ms} 
  G.~R.~Dvali, Q.~Shafi and R.~K.~Schaefer,
  Phys.\ Rev.\ Lett.\  {\bf 73}, 1886 (1994)
  [hep-ph/9406319].
\bibitem{Planck} 
  Planck Blue Book. 
  [http://www.rssd.esa.int/SA/PLANCK/docs/Bluebook-ESA-SCI{\%}282005
  {\%}291\_ V2.pdf]
\bibitem{GarciaBellido:1997wm} 
  J.~Garcia-Bellido and A.~D.~Linde,
  Phys.\ Rev.\ D {\bf 57}, 6075 (1998)
  [hep-ph/9711360].
\bibitem{BasteroGil:1999fz} 
  M.~Bastero-Gil, S.~F.~King and J.~Sanderson,
  Phys.\ Rev.\ D {\bf 60}, 103517 (1999)
  [hep-ph/9904315].
\bibitem{Zlatev:1997vd} 
  I.~Zlatev, G.~Huey and P.~J.~Steinhardt,
  Phys.\ Rev.\ D {\bf 57}, 2152 (1998)
  [astro-ph/9709006].
\bibitem{Felder:2001kt} 
  G.~N.~Felder, L.~Kofman and A.~D.~Linde,
  Phys.\ Rev.\ D {\bf 64}, 123517 (2001)
  [hep-th/0106179].
\bibitem{Felder:2000hq} 
  G.~N.~Felder and I.~Tkachev,
  Comput.\ Phys.\ Commun.\  {\bf 178}, 929 (2008)
  [hep-ph/0011159].
\bibitem{Frolov:2008hy} 
  A.~V.~Frolov,
  JCAP {\bf 0811}, 009 (2008)
  [arXiv:0809.4904 [hep-ph]].
\bibitem{Coleman:1973jx}
  S.~R.~Coleman and E.~J.~Weinberg,
  Phys.\ Rev.\ D {\bf 7} (1973) 1888.
\bibitem{Huq:1975ue} 
  M.~Huq,
  Phys.\ Rev.\ D {\bf 14}, 3548 (1976).
\bibitem{Larson:2010gs} 
  D.~Larson, J.~Dunkley, G.~Hinshaw, E.~Komatsu, M.~R.~Nolta, C.~L.~Bennett, B.~Gold and M.~Halpern {\it et al.},
  Astrophys.\ J.\ Suppl.\  {\bf 192}, 16 (2011)
  [arXiv:1001.4635 [astro-ph.CO]].
\bibitem{Kofman:1994rk}
  L.~Kofman, A.~D.~Linde and A.~A.~Starobinsky,
  Phys.\ Rev.\ Lett.\  {\bf 73} (1994) 3195
  [hep-th/9405187].
\bibitem{Felder:2000hj} 
  G.~N.~Felder, J.~Garcia-Bellido, P.~B.~Greene, L.~Kofman, A.~D.~Linde and I.~Tkachev,
  Phys.\ Rev.\ Lett.\  {\bf 87}, 011601 (2001)
  [hep-ph/0012142].
\bibitem{Enqvist:2004ey} 
  K.~Enqvist, A.~Jokinen, A.~Mazumdar, T.~Multamaki and A.~Vaihkonen,
  Phys.\ Rev.\ Lett.\  {\bf 94}, 161301 (2005)
  [astro-ph/0411394].
\bibitem{Barnaby:2006cq} 
  N.~Barnaby and J.~M.~Cline,
  Phys.\ Rev.\ D {\bf 73}, 106012 (2006)
  [astro-ph/0601481].
\bibitem{Barnaby:2006km} 
  N.~Barnaby and J.~M.~Cline,
  Phys.\ Rev.\ D {\bf 75}, 086004 (2007)
  [astro-ph/0611750].
\bibitem{Rehman:2009yj} 
  M.~U.~Rehman, Q.~Shafi and J.~R.~Wickman,
  Phys.\ Lett.\ B {\bf 688}, 75 (2010)
  [arXiv:0912.4737 [hep-ph]].

\bibitem{Mazumdar:2010sa} 
  A.~Mazumdar and J.~Rocher,
  Phys.\ Rept.\  {\bf 497}, 85 (2011)
  [arXiv:1001.0993 [hep-ph]].
\bibitem{Dufaux:2010cf} 
  J.~-F.~Dufaux, D.~G.~Figueroa and J.~Garcia-Bellido,
  Phys.\ Rev.\ D {\bf 82}, 083518 (2010)
  [arXiv:1006.0217 [astro-ph.CO]].
\bibitem{Frolov:2010sz} 
  A.~V.~Frolov,
  Class.\ Quant.\ Grav.\  {\bf 27}, 124006 (2010)
  [arXiv:1004.3559 [gr-qc]].
\bibitem{Chambers:2007se} 
  A.~Chambers and A.~Rajantie,
  Phys.\ Rev.\ Lett.\  {\bf 100}, 041302 (2008)
  [Erratum-ibid.\  {\bf 101}, 149903 (2008)]
  [arXiv:0710.4133 [astro-ph]].


  
     
  \end{thebibliography}
\end{document}